\documentstyle[11pt,citesort]{article}
\input epsf
\def \a {\alpha}
\def \b {\beta}

\def\th{\theta}

\def \I {{{}_{\! I}}}
\def \J {{{}_{\! J}}}
\def \is {\! & \! = \! & \! }
\newcommand{\newsubsection}[1]{
\vspace{1cm}
\pagebreak[3]
\addtocounter{subsection}{1}
\addcontentsline{toc}{subsection}{\protect
\numberline{
\arabic{subsection}}{#1}}
\noindent{\large\bf 
#1}
\nopagebreak
\vspace{2mm}
\nopagebreak}

\newcommand{\figuur}[3]{
\begin{figure}[t]\begin{center}
\leavevmode\hbox{\epsfxsize=#2 \epsffile{#1.eps}}\\[3mm]
\parbox{15.5cm}{\small
\it #3}
\end{center}
\end{figure}}
\newlength{\extraspace}
\newlength{\extraspaces}
\newcommand{\ba}{\begin{eqnarray}
\addtolength{\abovedisplayskip}{\extraspaces}
\addtolength{\belowdisplayskip}{\extraspaces}
\addtolength{\abovedisplayshortskip}{\extraspaces}
\addtolength{\belowdisplayshortskip}{\extraspaces}}
\newcommand{\ea}{\end{eqnarray}}
\newcommand{\be}{\begin{equation}
\addtolength{\abovedisplayskip}{\extraspaces}
\addtolength{\belowdisplayskip}{\extraspaces}
\addtolength{\abovedisplayshortskip}{\extraspace}
\addtolength{\belowdisplayshortskip}{\extraspace}}
\newcommand{\ee}{\end{equation}}

\begin{document}
\addtolength{\baselineskip}{.8mm}
\def\calN{{\cal N}}
\def\Box{\square}
\def \Tr{\mbox{Tr\,}}
\def \tr{\mbox{Tr\,}}
\def \a{\alpha}
\def \da{{\dot a}}
\def \db{{\dot b}}
\def \dc{{\dot c}}
\def \dd{{\dot d}}
\def \CH{\mathfrak {CH}}
\def \b{\beta}
\def \vt{\vartheta}
\def \ttheta{{\tilde \theta}}
\def \tSigma{{\tilde \Sigma}}
\def\ignorethis#1{}
\def\ar#1#2{\begin{array}{#1}#2\end{array}}
\def\bear{\begin{eqnarray}}
\def\eear{\end{eqnarray}}
\def\p{\partial }
\def\IR{{\bf R}}
\def\nul{{\mbox{\tiny $0$}}}
\def \two {{{}_{2}}}

\newcommand{\ggt}{\!>\!\!>\!}
\newcommand{\llt}{\!<\!\!<\!}
\def\dag{\dagger}
\def\bea{\begin{eqnarray}}
\def\eea{\end{eqnarray}}
\def\appendix#1{
  \addtocounter{section}{1}
  \setcounter{equation}{0}
  \renewcommand{\thesection}{\Alph{section}}
  \section*{Appendix \thesection\protect\indent \parbox[t]{11.15cm}
  {#1} }
  \addcontentsline{toc}{section}{Appendix \thesection\ \ \ #1}
  }

\begin{titlepage}
\begin{center}

{\hbox to\hsize{ \hfill PUPT-2050}} {\hbox to\hsize{ \hfill
hep-th/0209215}}

\bigskip

\vspace{6\baselineskip}

{\Large \bf Bit Strings from ${\cal N}$=4 Gauge Theory}
\bigskip
\bigskip
\bigskip

{\large Diana Vaman and Herman Verlinde}\\[1cm]

{ \it Physics Department, Princeton University, Princeton, NJ 08544}\\[5mm]

\vspace*{1.5cm}

{\bf Abstract}\\
\end{center}
\noindent We present an improvement of the interacting string bit
theory proposed in hep-th/0206059, designed to reproduce the
non-planar perturbative amplitudes between BMN operators in ${\cal
N}=4$ gauge theory. Our formalism incorporates the effect of
operator mixing and all non-planar corrections to the inner
product. We use supersymmetry to construct the bosonic matrix
elements of the light-cone Hamiltonian to all orders in $g_2$, and
make a detailed comparison with the non-planar amplitudes obtained
from gauge theory to order $g_\two^2$. We find a precise match.

\end{titlepage}

\newpage

\newsubsection{Introduction}

The correspondence \cite{bmn} between ${\cal N}\!=\! 4$
gauge theory at large R-charge and string theory
in a plane wave geometry \cite{bla} provides a promising new
example of a string/gauge theory duality, opening up a new road
for studying both sides of the duality.
The BMN dictionary, between single and
multi-trace operators and single and multi-string states, has by
now withstood many tests, but several puzzles remain.
In particular, it would be desirable to have a precise
characterization of the degrees of freedom within the ${\cal N}\! =\! 4$
theory that survive the BMN limit, and a comprehensive
formalism for describing their interactions.  Ideally, such a formalism
should allow for a clear comparison with the perturbative continuum
string field theory, as well as encompass all non-perturbative
D-brane configurations.

In \cite{bits}, an interacting string bit model \cite{thornbits}
was proposed with the aim of providing a useful interpolating
formalism between the gauge theory and the dual continuum string
theory. The idea is to design the string bit Hamiltonian such that
its matrix elements reproduce the perturbative gauge theory
amplitudes, including the non-planar corrections. Since the
appearance of \cite{bits}, however, more precise insights into the
structure of both the gauge theory and the bit string theory made
clear that the original proposal of \cite{bits} needs refinement.
Recent work by several groups \cite{bkpss}\cite{boston2}\cite{gmr}
(following up on earlier works
\cite{boston}\cite{kpss}\cite{bn}\cite{durham}\cite{huang}) has
produced a quite complete result for the leading non-planar
contributions to operator mixing coefficients \cite{bianchi} and
conformal dimensions for all bosonic two impurity BMN operators.
In this paper, we continue the program proposed in \cite{bits},
while incorporating these new results. Using the gauge theory and
supersymmetry as our guide, we construct the matrix elements of
the light-cone Hamiltonian to all orders in $g_2$. We compute the
order $g_\two^2$ matrix elements of $H$ between bosonic two
impurity states, and find a precise correspondence with the gauge
theory amplitudes.

\newsubsection{Free bit string theory}

We start with a short review of the bit string model proposed in
\cite{bits}. The bit strings consist of $J$ string bits, where in
the end one sends $J\! \to\! \infty$. To describe the motion of
the string bits, we introduce $J$ supersymmetric phase space
coordinates $\{ p_n^i, x_n^i, \theta_n^a, \tilde\theta_n^a\}$,
with $n=1, \ldots, J$, satisfying the usual canonical commutation
relations \be [\, p^i_n , x^j_m\, ]  = {\rm i} \delta^{ij} \delta_{mn} \,
, \qquad \ \ \{ \theta^a_n, \theta^b_m \}  = {1\over 2}
\delta^{ab} \delta_{mn} \, , \qquad \ \ \{ \tilde\theta^a_n,
\tilde\theta^b_m \}  = {1\over 2} \delta^{ab} \delta_{mn}. \ee
Since the string bits are assumed to be indistinguishable, we must
divide out the symmetric group $S_J$, acting via permutation on
the labels $n$.

Upon quantization, the Hilbert space splits up as a direct sum of
``twisted sectors'' ${\cal H}_\gamma$ labelled by conjugacy
classes $\gamma$ of the symmetric group $S_J$. On ${\cal
H}_\gamma$ we can act operators ${\cal O}(p,x,\theta)$ that are
left invariant under the action of the centralizer subgroup
$C_\gamma$ of $\gamma$: \be \label{inv} \qquad \{ \,p^i_n, x^i_n,
\theta^{a}_n\}\; \rightarrow \ \{\, p^i_{\sigma(n)},
x^i_{\sigma(n)}, \theta^{a}_{\sigma(n)}\}\, , \qquad \ \ \sigma
\in C_\gamma. \ee Since an arbitrary group element
$\gamma\in\,S_J$ is conjugate to a permutation of the form \be
\label{cycle} (J_1)(J_2)\ldots\,(J_s) \ee with $(J_i)$ a cyclic
permutation of length $J_i$,  we thus obtain a sum over
multi-string Hilbert spaces, each string with a discretized
worldsheet consisting of $J_\ell$ bits with $\sum_\ell J_\ell =J$.

The invariance (\ref{inv}) under the stabilizer subgroup imposes
the constraint $(\, U_\ell)^{J_\ell} = 1$ on each string, with
$U_\ell$ the operator that translates the string bits by one unit
on the $\ell$-th string. The ``overall'' translation operator $U =
\otimes_{\strut \! \! \ell} \, U_\ell$ is defined to act via \be U
X_n U^{-1} =X_{\gamma(n)} \ee with $X_n = \{ p_n^i, x_n^i,
\theta_n^a\}$. Correspondingly, the free light-cone supersymmetry
generators and Hamiltonian involve ``hopping terms'' that depend
on the choice of twisted sector $\gamma$: \be \label{expa}
Q_{\nul,1} = Q^{{}^{\! (0)}} \! + \lambda \, Q^{{}^{\! (1)}} \ ,
\qquad \ \ Q_{\nul, 2} = \tilde{Q}^{{}^{(0)}} - \lambda
\tilde{Q}^{{}^{(1)}}\, , \qquad \ \ H_\nul = H^{{}^{\! (0)}} \! +
\lambda \, H^{{}^{\! (1)}} \! + \lambda^2 H^{{}^{\! (2)}}\, , \ee
with \ba \label{lcgen} Q^{{}^{\! (0)}}\!
\is \! \sum_n (p_{n}^{\, i} {\gamma}_i
\, \theta_n \! -  x^{i}_n({\gamma}_i
\Pi)
\tilde\theta_n) \, , \qquad \qquad Q^{{}^{\! (1)}} \! =   \sum_n
(x^i_{\gamma(n)}\!\! - x^i_{n})
{\gamma}_i
\, \theta_n  \qquad 
\\[3.5mm]
H^{{}^{\! (0)}} \! \! \is \! \sum_n (  \frac{1}{2 } (p_{i,n}^2 +
 x_{i,n}^2) + 2{\rm i}\, \tilde\theta_n\Pi
\theta_n) \, , \\[3.5mm]
H^{{}^{\! (1)}}\!\! \is\! - \sum_n {\rm i}(\theta_n
\th_{\gamma(n)} -\tilde\theta_n\tilde\theta_{\gamma(n)} ) \, ,
\qquad \quad \ \ \ H^{{}^{\! (2)}}\! =  \sum_n \frac{1}{2 }
(x^{i}_{\gamma(n)} \!\! - x^i_n)^2 \, .\ea These expressions are
the most straightforward discretization of the supercharges of the
continuum string theory given in \cite{mets}. It is easy to verify
that they generate a closed supersymmetry algebra. Note further
that the hopping terms are defined such that they just act within
each separate string.

To proceed, it will be useful to introduce creation and
annihilation operators for the individual string bits via \ba
x_{m}^{i}\! \is \frac{1}{\sqrt{2}}(a_{m}^{i}\, + a_{m}^{i\ \dag})
\qquad \qquad \; {p_{m}^{i}}\, =\, \frac{{\rm i}}{\sqrt{2}}(a_{m}^{i}\,
-
a_{m}^{i\ \dag}) \nonumber \\[2.5mm]
\theta_n \! \is \! \frac{1}{2}\left( \beta_n+\beta_n{}^\dag\right)
\qquad \qquad \Pi\theta_n=\frac{i}{2}\left(
\tilde{\beta}_n-\tilde{\beta}_n{}^\dag\right)
\\[2.5mm]
\tilde\theta_n\! \is \! \frac{1}{ 2}\left(\tilde \beta_n{}
+\tilde\beta^\dag_n\right) \qquad \qquad
\tilde\theta_n\Pi=\frac{i}{ 2}\left(
\beta_n{}-\beta^\dag_n\right)\, . \nonumber \eea This choice of
fermionic creation and annihilation operators breaks the $SO(8)$
rotation symmetry down to $SO(4)\times SO(4)$.

We can now define the untwisted vacuum state $|0\rangle$ as the
simple tensor product of the vacuum states of the individual
string bits \be \label{vac} a_{m}^{i} |0\rangle =0\, , \ \ \qquad
\ \beta_{m}^{\da}|0\rangle =0\, , \ \qquad \
\tilde\beta_m^\da|0\rangle = 0 \, \qquad \langle 0|0\rangle =1\, .
\ee This state describes $J$ disconnected string bits, i.e. $J$
separate short strings of unit length,  all in their ground state.
We wish define operators that, when acting on $|0\rangle$ produce
the ground state of a collection of long strings of length
$J_\ell$ with $\sum_\ell J_\ell =J$.

For all elements $\sigma \in S_J$, we associate a corresponding
twist operator $\Sigma_\sigma$ that implements the permutation
$\sigma$ on the string bits:
\be \label{sigma1}
\qquad \qquad \Sigma_{\sigma} X_n = X_{\sigma(n)} \,
\Sigma_{\sigma}. \ee This operator acts non-trivially on the
twisted sectors via \be \label{sigma2} \Sigma_{\sigma}: \quad
{\cal H}_\gamma \rightarrow {\cal H}_{\gamma \circ \sigma}\, . \ee
In particular, we can define the twisted vacuum states
$|\gamma\rangle \in {\cal H}_\gamma$ as follows \cite{zhou} \be
|\gamma \rangle = {1 \over \sqrt{N_\gamma}} \sum_{\sigma \in
\gamma} \Sigma_\sigma \, |0\rangle\, . \ee Here $N_\gamma$ is the
number of elements in the conjugacy class $\gamma$. This twisted
vacuum state also satisfies \be a_{m}^{i}\,|\gamma\rangle =0\, , \
\ \qquad \ \beta_{m}^{\da}\,|\gamma\rangle=0\, , \ \ \qquad \
\tilde\beta_{m}^{\da}\,|\gamma\rangle=0\, , \ \ \qquad \
\langle\gamma|\gamma\rangle =1. \ee This ground state
$|\gamma\rangle$ describes a collection of strings in their ground
state,  of length $J_i$ depending on the decomposition
(\ref{cycle}) of $\gamma$ into cyclic permutations. In particular,
the single string state corresponds to the long cycle
$\gamma_1=(1~2 \ldots J)$, and double string states to
permutations of the form $\gamma_2 = (1~2 \ldots J_1)(J_1 \ldots
J)$.

\newsubsection{BMN dictionary}

BMN proposed a concrete dictionary between operators in the string
bit theory and operators of large R charge in ${\cal N}\!=\! 4$
gauge theory. This correspondence is based on the identification
of the string light-cone Hamiltonian with $H=\Delta - J$, where
$J$ (the total number of string bits) equals the total R charge.
The hopping parameter $\lambda$ in the string bit Hamiltonian gets
identified with the 't Hooft coupling of the gauge theory \be
\lambda^2 = \frac{g^{2}_{YM}\,N}{8\pi^2}. \ee The free string bit
theory then describes the planar limit $N\! \to\! \infty$. Our
goal is to construct the effective string interactions that arise
for $g_\two = J^2/N$ finite.

\newcommand{\one}{\mbox{\small $1$}}

In \cite{bkpss}\cite{boston2}\cite{gmr} a complete analysis was
given of the leading order non-planar corrections to the conformal
dimension of a specific class of BMN operators. We will test our
formalism by showing that it reproduces the same results. The
normalized one-string BMN operators considered are of the form
\be
O_{p}^{J}=\frac{1}{\sqrt{JN^{J+2}}}\sum_{l=1}^{J}e^{2\pi\,ipl/J}
{\rm Tr}(\phi\,Z^{l}\psi\,Z^{J-l}). \ee To write the corresponding
bit string state we introduce the operator \cite{zhou}
\be
O_{p,\gamma_1}^{J}=\frac{1}{J}\Big(\sum_{k=1}^{J}a_{\gamma_{1}(k)}^{\dag}
e^{-2\pi\,ipk/J}\Big) \Big(\sum_{l=1}^{J}b_{\gamma_{1}(l)}^{\dag}
e^{2\pi\,ipl/J}\Big) \label{o1} \ee where $\gamma_1$ is
the long cycle of length $J$. The associated state is obtained by
acting on the single string vacuum state $|\gamma_1 \rangle$, and
then summing over all conjugations of $\gamma_1$: \be
\label{steen} |O_{p}^{J}\rangle= {\one \over {\mbox{\small
$\sqrt{J! J }$}}}\sum_{\tilde{\gamma}_1 = h^{{}^{\! -1}} \!
\gamma_1 h} O_{p,\tilde\gamma_1}^{J} |\tilde{\gamma}_1\rangle\, ,
\label{pstate} \ee with $h\in S_J$. This state is normalized to
have unit norm.

Similarly, we can construct a normalized two-string state
corresponding to the normalized double trace BMN operator \be
\frac{1}{\mbox{\small$\sqrt{J_{1}(J\!-\!J_1)
\,N^{J+2}}$}}\sum_{l=1}^{J_1}e^{2\pi\,ikl/{J_{1}}} {\rm
Tr}(\phi\,Z^{l}\psi\,Z^{J_{1}\,-l})\, {\rm Tr} (Z^{J-J_1}). \ee
via
\be \label{stwee} |O_{k}^{J_1}\rangle={\one\over \mbox{\small $
\sqrt{J!J_1(J\!-\!J_1)}$}} \sum_{\tilde{\gamma}_2 = h^{{}^{\! -1}}
\! \gamma_2 h} O_{k,\tilde\gamma_2}^{J_1} |\tilde{\gamma}_2\rangle
\label{o2} \ee with \be O_{k,\gamma_2}^{J_1} =
\frac{1}{J_1}\Big(\sum_{l=1}^{J_1}a_{\gamma_{2}(l)}^{\dag}
e^{-2\pi\,ikl/{J_1}}\Big)
\Big(\sum_{l'=1}^{J_1}b_{\gamma_{2}(l')}^{\dag}
e^{2\pi\,ikl'/{J_1}}\Big) \ee where $\gamma_2$ is decomposed as
$(1~2 \ldots J_1)(J_1\!+1 \ldots J)$. Finally, the other type of
two-string state corresponding to \be \frac{1}{N^{J+2}}
\Tr(\phi\,Z^{J_{1}})\,\Tr(\psi\,Z^{J-J_{1}}) \ee is \be
|O_{0}^{J_{1}\,J_{2}}\rangle = {\one \over \mbox{\small $ \sqrt{J!
J_{1}\,(J\!-\!J_{1})}$}}\sum_{\tilde{\gamma}_2 = h^{{}^{\! -1}} \!
\gamma_2 h} O_{0,\tilde\gamma_2}^{J_1J_2} |\tilde{\gamma}_2\rangle
\ee with \be O_{0,\gamma_2}^{J_1J_2} \, = \, {\one \over
\mbox{\small $ \sqrt{J_{1}\,(J\!-\!J_{1})}$}}
\Big(\sum_{l=1}^{J_1}a_{\gamma_{2}(l)}^{\dag}\Big)
\Big(\sum_{l'=J_{1}\,+1}^{J}b_{\gamma_{2}(l')}^{\dag}\Big)
\label{o3} \ee

In the following we will often use the notation \be |1,p\rangle =
| O_p^J\rangle \, , \qquad \quad |2,k,y\rangle =
|O_{k}^{J_1}\rangle\, , \qquad \quad | 2, y \rangle =
|O_{0}^{J_{1}\,J_{2}}\rangle, \ee with $y= J_1/J$ a sub-unitary
parameter that parametrizes the relative length of the two
strings. These three states are all eigenstates of the free
Hamiltonian $H_\nul$ with respective eigenvalues equal to $E_p = 2
+ \lambda' \, p^2$, $ E_k = 2 + \lambda' k^2/y^2$ and $2$, with
\be \lambda' = {8\pi^2 \lambda^2\over  J^2}\, .\ee

Two final comments: (1) Note that the definition (\ref{vac}) of
the bit string vacuum state is uniquely selected by requiring that
is should correspond to the BPS operator $\Tr(Z^J)$: both are the
lowest energy eigen states. (2) Above we have made a direct
correspondence between the single and double trace BMN operators
and one and two string states. As we will see shortly, this
identification is in fact somewhat premature, since upon turning
on the effective string coupling \be g_\two = \frac{J^2} N \ee
single and multiple trace operators will inevitably start to mix.
For now, however, we will adopt the above direct identification
between the BMN operators and string bit states, leaving the
discussion of possible redefinitions to the concluding section.
First we wish to determine the form of the bit string
interactions, using the gauge theory as our guide.

\newsubsection{Inner Product at Finite $g_\two$}

A characteristic aspect of the gauge theory is that, even at zero
't Hooft coupling $\lambda$, the overlap between the BMN operators
has a non-trivial expansion in terms of $g_\two = J^2/N$, because
free Wick contractions can still generate a sum over non-planar
diagrams. In particular, there is a non-vanishing overlap between
single trace and multi-trace BMN operators
\cite{kpss,boston,durham}.

Since the $\lambda\! =\! 0$ theory is free, this structure can be
explicitly worked out by keeping track of the permutations $\sigma
\in S_J$ encoded in the Wick contractions between the $J$ string
bits of the ``in'' and ``out'' operators \cite{boston}. Since our
goal is to construct the bit string model in such a way that it
reproduces the gauge theory amplitudes, we need to incorporate
this structure by means of an appropriate choice of inner product.
Luckily, permutations of the string bits are already a natural
part of the story.

Recall that any permutation $\sigma$ can be factorized into a
product of simple permutations of the form $(nm)$. Let $h(\sigma)$
be the minimal number of simple permutations needed in this
factorization of $\sigma$. The inner product, that realizes the
combinatorics of the free gauge theory amplitudes in the string
bit language, is of the form \be \label{ip} \langle \, \psi_1 | \,
\psi_2 \rangle_{g_2} = \, \langle \, \psi_1 | \, S \, | \, \psi_2
\rangle_{\nul} \ee where $S$ is the following weighted sum over
all possible permutation operators \be \label{es} S =
\sum_{\sigma} {N^{-2 h(\sigma)}} \; \Sigma_\sigma. \ee

To understand the structure of this inner product, let us consider
the first few terms in the expansion (\ref{es}) a bit more
closely. Writing \be \label{sexp} S = 1 + {1\over N} \Sigma_2 +
{1\over N^2} \, \Sigma_3 + \ldots \ee we find for the first order
term \be \Sigma_2 = \sum_{n<m} \Sigma_{(nm)}\, ,
\ee with $(nm)$
the simple permutation of order 2. This operator $\Sigma_2$
represents a basic cubic string joining and splitting interaction.
The special role of $\Sigma_2$ will become more apparent in the
following.

The second order term $\Sigma_3$ is \be \Sigma_3 = \sum_{m<n \atop
{m<k<l \atop k,l \neq n}} \Sigma_{(mn)(kl)} + \sum_{m<n<k}
(\Sigma_{(mnk)} + \Sigma_{(knm)}). \ee When acting on a single
string state, it can either split the string into three separate
strings, or induce a subsequent splitting and joining, producing a
new reordered  single string state. It is straightforward to
verify that the Feynman diagram produced by the corresponding free
Wick contraction between single trace ``in'' and ``out'' BMN
operators has genus 1. Now, since $(mnk) = (mk)(kn)$, we see that
\be \label{sigmasq} \Sigma_3 = {1\over 2} \Bigl((\Sigma_2)^2 -
\mbox{\small $ J(J\! -\! 1)$}\Bigr). \ee The $c$-number term
becomes negligible in the limit of large $J$, since $\Sigma_3$ in
(\ref{sexp}) comes with a prefactor of $1/N^2$. The physical
meaning of the identity (\ref{sigmasq}) is that all second order
string interactions can be thought of as the result of two
elementary string interactions. Generalizing this observation to
higher orders, it is natural to suspect that in the the limit of
large $J$, we can write \be \label{es2}
 S = e^{g_2 \Sigma} \, ,
\qquad \qquad \ \ \Sigma \equiv \, {1\over
J^2} \,\Sigma_2 \, .
\ee We claim that the inner product (\ref{ip}) with (\ref{es2})
indeed corresponds to that of the free gauge theory at large $J$
but finite $g_\two$.

As a specific check, let us compute the genus $h$ contribution to
the overlap between two single string vacuum states \be
\label{4hgon} { 1\over (2h)!} \, \langle \gamma_1 |
(\Sigma_2)^{2h} | \gamma_1 \rangle. \ee This contribution is equal
to the total number of products of $2h$ simple permutations that,
when acting on a long cycle (single string) produce another long
cycle. This number can be evaluated as follows (see e.g. the
discussion in \cite{boston}): The product $\Sigma_2^{2h}$ involves
a sum over $2h$ pairs of positions, which (absorbing the factor
$1/(2h)!$) can be assumed to be ordered. The $2h$ bit pairs split
up the $J$ bits into $4h$ groups. Placing the $J$ string bits
along a circle, this produces a $4h$-gon, on which the $2h$ pairs
represent a specific gluing rule: each two corner points of the
$4h$-gon connected by a simple permutation must be glued together.
This gluing rule reflects the correspondence between simple
permutations and elementary string splitting or joining
interactions. The condition that the product of simple
permutations maps a long cycle to another long cycle, now
translates into the condition that the gluing produces a surface
of genus $h$. Via this reasoning, one obtains that (\ref{4hgon})
equals the number of ways of dividing $J$ bits into $4h$ groups
(which for large $J$ equals $J^{4h}/{4h}!$) times the number of
ways of gluing a $4h$-gon into a genus $h$ surface (which is
known to be equal to $(4h-1)!!\over {2h+1}$). So our bit string
inner product indeed reproduces the gauge theory result \be
 \langle \gamma_1 | S\,
| \gamma_1 \rangle = \sum_{h=0}^\infty {1\over (2h+1)!} \Bigl(
{g_\two \over 2}\Bigr)^{2h} = {2\over g_\two} \sinh (g_\two/ 2)
\ee As a further concrete check on the above reasoning, we have
explicitly worked out the genus 1 and 2 contributions in Appendix
B.

Another check on our inner product is obtained by considering the
matrix elements of the first and second order terms in $S$ between
the special class of states introduced earlier. A straightforward
calculation (along the lines of \cite{zhou}) shows that a single
action of $\Sigma$ produces the same non-zero ``three point
functions'' between the normalized two-impurity states as those
obtained in the free gauge theory
\cite{kpss}\cite{durham}\cite{boston} \ba \label{cijk} C_{pk y} \!
& \equiv & \! \langle\, 2, k , y| \, \Sigma\, | \,1, p\, \rangle\,
= \, \sqrt{1-y \over Jy} {\sin^2 (\pi p y) \over
\pi^2 (p- k/y)^2} \, , \nonumber \\[3mm]
C_{p y} \! & = &  \langle 2, y \, | \, \Sigma\, | \, 1, p\,
\rangle \ \ =\, - {\sin^2 (\pi p y)\over \sqrt{J} \pi^2 p^2} \,
.\ea
{}From the definition of $\Sigma$, it is furthermore clear that
these ``three point functions'' form a complete set in the sense
that \be \label{ope} \Sigma \, | 1, p \rangle = \sum_{k,y} C_{pky}
\, |2, k,y\rangle\;  + \; \sum_y C_{py} |2,y\rangle \, . \ee

As a technical aside, we note that the above decomposition
relation in fact reveals a conceptual subtlety, which was noted in
\cite{zhou}. Namely, the sum over $k$ in (\ref{ope}), strictly
speaking, must be extended to include values of order $J_1$. In
this regime, however, one can no longer make the approximation
$\sin^2(\pi(py-k)/J_1) \simeq \pi^2 (py-k)^2/J_1^2$ that was used
to derive (\ref{cijk}). Problems of this sort often arise in
discretized models, and we will deal with it in the usual manner:
we will simply truncate the Hilbert space of the bit string model
to include only those frequencies $k$ negligibly small compared to
$J$. This restriction should become insignificant upon taking the
large $J$ limit.

The second order matrix element of $S$ between the single string
states, representing the one-loop contribution due to successive
splitting and joining, can be similarly be obtained, either via
direct computation, or by using factorization \be \label{adef}
A_{pq} \, \equiv \, \frac 12\langle\, 1, q | \Sigma^2 | \, 1,
p\,\rangle\, = \, {1\over 2} \Bigl( \sum_{k,y} C_{pky} C_{qky} +
\sum_y C_{py}C_{qy}\Bigr)\, . \ee This
relation 
matches the factorization property of the inner product of the
free gauge theory, which was first derived in \cite{huang}. The
explicit form of $A_{pq}$ is as given in \cite{kpss}
\cite{boston}.

Finally, we need to emphasize that the above amplitudes do not yet
represent proper string interactions. String interactions are
associated with non-trivial matrix elements of the light-cone
Hamiltonian. The $g_\two$-dependence of the inner product can
obviously be transformed away by a redefinition of the single and
multi-string states (see the concluding section). For now we will
stick to the above basis, so that the relation to the gauge theory
is most apparent.

\newcommand{\gtt}{{{}^{{}_ {>}}}}
\newcommand{\stt}{{{}^{{}_{<}}}}
\newcommand{\gst}{{{}^{{}_{\geq}}}}
\newcommand{\sgt}{{{}^{{}_{\leq}}}}
\newcommand{\gt}{{{}^{ >}}}
\newcommand{\st}{{{}^{<}}}

\newcommand{\widetild}{}

\newsubsection{Interactions and Supersymmetry}

The modification (\ref{ip}) of the inner product at finite string
coupling $g_\two$ indicates that we must also add new interaction
terms to the supersymmetry generators and Hamiltonian. Matrix
elements of the Hamiltonian at non-zero $g_\two$ can be expressed
in terms of the bare inner product (the one at $g_\two\! =\! 0$)
via \be \langle \, \psi_2| H\, | \, \psi_1 \rangle_{g_\two} \, =\,
\langle \, \psi_2| S H\, | \, \psi_1 \rangle_\nul\, . \ee
Hermiticity of $H$ thus requires that \be H = H^\dagger = S^{-1}
H^{\dag 0} S \ee where $H^{\dag 0}$ denotes the hermitian
conjugate relative to the bare inner product. A similar condition
holds for the supersymmetry generators.

Another non-trivial consistency requirement is the closure of the
interacting light-cone supersymmetry algebra (here {\footnotesize{
$I,J =1,2$}}  -- see eqn (\ref{expa})) \be \label{lcsym}
\delta^{{}^{IJ}}\! \{ \widetild{Q}_\I^\da, {Q}_\J^\db\}
=\;   \delta^{\da\db}  H \, +\, J^{\da\db}\, ,
\ee where $J^{\da\db}$ is a suitable contraction of gamma matrices
with the $SO(4) \times SO(4)$ Lorentz generators $J^{ij}$, see
\cite{mets}.

In this section we will write a new Ansatz for $Q^\da$ and $H$,
that will be hermitian relative to the new inner product, and will
produce non-trivial string interactions proportional to $g_\two$.
Our Ansatz will generate the light-cone supersymmetry algebra
(\ref{lcsym}), but only at the linearized level in the fermions,
that is, when inserted between string states with only bosonic
excitations (or between a purely bosonic and a fermionic one). In
principle, it should be straightforward to correct our Ansatz for
$Q$ by means of non-linear fermionic terms, so that the algebra
closes for all fermionic states as well. Our main interest in the
following, however, will be to compare our model with the gauge
theory computations, which so far have been done for bosonic
states. Because we expect that the non-linear fermionic correction
terms in the end will not modify these bosonic amplitudes, we will
leave their study to a future work.


To write the interacting generators, we will use the
correspondence with the gauge theory as our guide. The basic idea
will be the following. We will assume that the free supersymmetry
generators can be split into two terms \be \label{split} Q_\nul =
Q_\nul^\gt + Q_\nul^\st\, .
\ee
such that, in the interacting theory, $Q_\nul^\st$ 
will receive correction terms that induce string splitting and
joining {\it only} when acting on states to the right, while $Q_\nul^\gt$ 
will induce string splitting and joining {\it only} when acting on
states to the left. The underlying motivation for this assumption
is that, in the correspondence with the gauge theory, $Q_\nul^\st$
represents an interaction term of the form $\Tr(\theta [\bar Z,
\bar \phi])$ which naturally acts via a {\it double} Wick
contraction on the ``in'' BMN state, while only with a {\it
single} contraction on the ``out'' state. The double Wick
contraction can split up a trace, or join a product of two traces
into a single trace, while a single contraction can not.

Let us give an explicit example. The Wick contraction between
\begin{equation}
Q^\st = \Tr\left(\theta [\overline{Z},\overline{\phi}]\right) \,
\qquad {\rm and} \qquad  O = \sum_{l=0}^{J} q^l \Tr\left(\phi Z^l
\psi Z^{J-l}\right)
\end{equation}
with $q = e^{2\pi i p/J}$ has been found to be equal to
\cite{boston}
\bea
\label{danr}
Q^\st O \is -iN(q\! -\! 1)\sum_{l=0}^{J-1}q^l(\theta Z^l\psi
Z^{J-l-1})
\nonumber \\[2.5mm] && -i\frac{q}{q-1}\sum_{J_1=1}^{J-1}(\theta
Z^{J_1})(Z^{J-J_1-1}\psi) (1\!+\! q^{-1}\!\! -\! q^{J_1}\!\! -\!
q^{-J_1-1})
\\[2.5mm]
&& -\; i\sum_{J_1=1}^{J-1}\sum_{l=J_1+1}^{J}q^l(1-q^{-J_1-1})
(Z^m)(\theta Z^{l-J_1-1}\psi Z^{J-l})\, .\nonumber \label{nonnear}
\eea In string bit language, the single trace term corresponds to
the free action of the supercharge,  while the double trace terms
are due to an interaction term in $Q^\st$ proportional to
$g_\two$, that induces a single string splitting. In contrast, the
Wick contraction between \be \widetild{Q}^\gt =
\Tr\left(\overline{\theta} [{Z},{\phi}]\right) \, \qquad {\rm and}
\qquad O' = \sum_{l=0}^{J} q^l \Tr\left(\theta Z^l \psi
Z^{J-l}\right)
\ee
is simply \be \widetild{Q}^\gt O' \, = \, -iN
\sum_{l=0}^{J-1}q^l([Z, \phi] Z^l\psi Z^{J-l-1})\, . \ee Hence
this term $\widetild{Q}^{\gt}$, which is the hermitian conjugate
of $Q^\st$, acts just like the free supercharge.

We wish to incorporate this same structure into the definition of
the supersymmetry generators of the bit string theory. The above
two gauge theory calculations suggest that the division
(\ref{split}) should be made such that terms of the form
$\beta^\dag_m a_n$ are part of $Q_\nul^\st$ 
and will receive interaction terms
proportional to $g_\two$ when acting to the right, while all terms
of the form $\beta_m a_n^\dag$ are part of $Q_\nul^\gt$ 
and remain free when acting to the right.

With this motivation, we will now adopt the following Ansatz for
the supersymmetry generators for finite $g_\two$ \be \label{nsusy}
 Q = Q_\nul^\gt +
 S^{-1} Q_\nul^\st S\, ,
\ee where the $>$ superscript indicates the terms that contain
fermionic annihilation operators $\beta_m$ only, while $<$ denotes
terms with only $\beta^\dag_m$'s. In particular, \be \label{q12}
Q^{{}^{(1)}}{}^\gtt =-\sum_n (x^i_{\gamma(n)}-x^i_n)\gamma^i
\beta_n~~~~;~~~~
Q^{{}^{(1)}}{}^\stt=-\sum_n(x^i_{\gamma(n)}-x^i_n)\gamma^i
\beta_n{}^\dag. \ee The Ansatz (\ref{nsusy}) by construction
satisfies the hermiticity condition $Q^\dagger = \widetild{Q}$,
relative to the new inner product.


A priori, since $S$ contains terms of arbitrarily high powers in
$g_\two$, the new supersymmetry generators ${Q}$ in (\ref{nsusy}) 
appear to have an infinite $g_\two$ expansion. The gauge theory
supercharges, on the other hand, can effectuate (if we assume they
need at least one Wick contraction with either the ``in'' our
``out'' BMN state) at most a single string splitting or joining
interaction. It indeed turns out that, also in our bit string
model, only a linear interaction term survives
\be \label{nnsusy} Q = Q_\nul + g_\two [\, Q_\nul^\st , \Sigma\, ]
\, ,
\ee provided we take the strict large $J$ limit. This
simplification of the Ansatz (\ref{nsusy}) follows from the fact
that in this limit \be \label{double} [ [\, Q^\st_\nul,\Sigma\,
],\Sigma\, ] = 0 \,.
\ee To derive this identity, we note that the double commutator
with $\Sigma$ can be reduced to a triple (rather than quadruple)
summation over the $J$ sites, since the indices in the simple
permutations in the two $\Sigma$ factors have to coincide, or
differ by at most one unit, in order to give a non-zero result
(for finite $J$). This triple summation is insufficient to
overcome the $1/J^4$ pre-factor, and thus the double commutator
(\ref{double}) vanishes in the strict large $J$ limit.

It is straightforward to obtain an explicit form of the
interaction term in (\ref{nnsusy}), by letting it act on an
arbitrary state $|\psi_{\gamma_1}\rangle$ in a twisted sector
${\cal H}_\gamma$: \ba [\, Q^\stt_\nul,\Sigma\, ]
|\psi_{\gamma_1}\rangle \is {1\over J^2} \sum_{{m<n}}\, [\,
Q^\stt_\nul , \Sigma_{mn}\, ]\, |\psi_{\gamma_1}\rangle \; =\;
{1\over J^2}
\sum_{m<n}\Sigma_{mn}\Bigl(Q^\stt_{\gamma_2}-Q^\stt_{\gamma_1}\Bigr)
|\psi_{\gamma_1}\rangle \eea where $\gamma_2 = \gamma_1 \circ
(mn)$. Inserting the explicit form (\ref{q12}) of the
supersymmetry generator gives \ba \label{expli} [\, Q^\stt_\nul,
\Sigma \, ]\is {\lambda \over 2 J^2} \sum_{m,n} \Sigma_{mn}\bigg(
(x^i_{\gamma(m)}\!\!- \!x^i_{\gamma(n)})\gamma_i\, \beta_m^{\,
\dag}   + (x^i_m\! -\! x^i_n)\gamma_i\, \beta^{\,
\dag}{}_{\hspace{-6pt}\gamma^{{}^{\! -1}}\! (m)}
\qquad  \\[2.5mm]\nonumber
& & \qquad \qquad \qquad \qquad \qquad   + \; \delta_{n\gamma(m)}
\Bigl(x^i_n \gamma_i \beta_m^{\, \dag}  \!-x^i_m \gamma_i \beta^{\
\dag}_n \Bigr)\bigg) \, .\nonumber \ea This interaction term has a
finite strength, because it is involves a double sum
over $J$ sites, which compensates for the $1/J^2$ factor in front.

Let us now verify that the above interaction term, when acting on
the two impurity single string state $|O_p^J\rangle$, produces the
same result (\ref{danr}) as found the gauge theory. We can choose
the one string state $|O_p^J\rangle$ to lie in the twisted sector
labelled by $\gamma_1=(1 ~2\dots J)$. Acting with the interaction
term of the supercharge then produces\footnote{Here, in order to
compare with the gauge theory calculation leading to (\ref{danr}),
we keep in $Q$ only the term with $a^i$ annihilators, leaving out
the creation and $b^i$ modes. The $b^i$ terms would correspond in
the gauge theory to a term of the form $\Tr
\theta[\bar\psi,\bar{Z}]$.} \ba {1\over J^2}\sum_{m<n}
\Sigma_{mn}\, (\, Q^\stt_{\gamma_2}\! -Q^\stt_{\gamma_1} ) |O_p^J
\rangle\is\! {\lambda \over J} \sum_{J_1} \Sigma_{JJ_1}\Bigl[
(a^i_J\! -a_{J_1}^i )\gamma^i(\beta_{J_1-1}^\dag \!\! -
\beta_{J-1}^\dag) \Bigr] \sum_{k,l=1}^J a_k^\dag b_l^\dag
q^{l-k}\, |\gamma_1 \rangle\nonumber \eea
with $q= e^{-{2\pi i p\over J}}$.
Decomposing the sum over
the position of the impurity $b_l^\dag$ as \bea
\sum_{l=0}^{J_1-1} b_l^\dag q^l 
&\equiv& A~~~~~~~~~ \sum_{l=J_1}^{J-1} b_l^\dag q^l \equiv B \eea
the action of $[Q^{\stt},\Sigma_{J_1J}]$ on the one-string state
is given by \be [Q^{\stt},\Sigma_{J_1J}]|O_p^J\rangle =
\Bigl(1-q^{-J_1}\Bigr) \left[\Bigl(\beta_{J_1-1}^\dag B -
\beta_{J-1}^\dag A\Bigr) +\Bigl(-\beta_{J-1}^\dag B
+\beta_{J_1-1}^\dag A\Bigr)\right]\, |\gamma_2 \rangle \label{qs}
\ee where the first two terms on the right-hand side of (\ref{qs})
describe states with a fermionic impurity on a string of length
$J_1$ respectively $J\!-\! J_1$ and a bosonic impurity on the
complementary string,
 while the last two terms in (\ref{qs})
are states with both impurities sitting on the same string of
length $J_1$ and respectively $J-J_1$. Given the fact that one is
supposed to sum over all cyclic permutations, the states where the
bosonic impurity is placed on a different string than the
fermionic one \footnote{Just like in the gauge theory, a one-string state
of length $J$ and momentum $p$ with a single impurity vanishes for
non-zero $p$, due to the imposed
invariance under cyclic permutations of $\gamma_1$.}
yield \be \Bigl(1-q^{-J_1}\Bigr) \bigg(\beta_{J_1-1}^\dag
b^\dag_{J-1} \sum_{l=J_1}^{J-1}q^l \bigg)|\gamma_1 \rangle\, =
\frac{1}{q-1} \Bigl(2-q^{-J_1} -q^{J_1}\Bigr) \beta_{J_1-1}^\dag
b^\dag_{J-1} |\, \gamma_2 \, \rangle \label{diff} \ee which by the
dictionary we have established between the string bit and the
gauge theory is equal to \be\frac{1}{q-1} \Bigl(2-q^{-J_1}
-q^{J_1}\Bigr) {\rm Tr}(Z^{J_1-1}\theta) {\rm
Tr}(Z^{J-J_1-1}\Psi).\label{gdiff} \ee The states where the
impurities sit on the same string are \be \Bigl(1-q^{-J_1}\Bigr)\,
\beta^\dag_{J-1} \sum_{l=J_1}^{J-1} b^\dag_l q^l|\, \gamma_2 \,
\rangle\label{same} \ee and the same dictionary relates them to a
gauge theory operator \be\Bigl(1-q^{-J_1}\bigg) \sum_{l=J_1}^{J-1}
q^l\, {\rm Tr}(Z^{J-1})\, {\rm Tr}(Z^{l-J_1}\Psi
Z^{J-2-l}\theta).\label{gsame} \ee

Finally, note that out the four possible configurations of
(\ref{qs}) we have discussed only two, with the bosonic impurity
placed on the string of length $J\!-\! J_1$. The other two are
taken into account as we sum over $J_1$: they appear when $J_1$
equals $J_2=J\! -\! J_1$ with the bosonic impurity placed on the
string of length $J\! -\! J_2=J_1$. Thus in summing over $J_1$ we
find that $[Q^{\stt},\Sigma_{J_1J}]$ is given by twice
(\ref{diff}) and (\ref{same}).  Comparing the string bit
calculation with the gauge theory we see that (\ref{gsame})
reproduces the last term in (\ref{danr}) while (\ref{gdiff})
corresponds to the second term in (\ref{danr}). There are subtle
differences, which however disappear in the large $J$ limit, due
to the fact that the action
of $\Tr(\theta [Z,\phi])$ in the gauge theory 
doesn't conserve $R$-charge: it reduces or increases the number of
string bits $J$ by one unit. The supersymmetry generators in the
string bit picture, on the other hand, preserve the total length
of the bit strings.

\newsubsection{Matrix Elements of the Hamiltonian} 

In this section we will evaluate the matrix elements of the string
bit Hamiltonian, and compare it with the ones computed in gauge
theory. So far, all gauge theory computations have been done for
BMN states with two bosonic impurities.

Starting from our Ansatz (\ref{nsusy}) for the supercharges, we
can now define the matrix elements of $H$ via \be \label{hdiag}
\langle \psi_2 | (\delta^{\da\db} H + J^{\da\db})| \psi_1
\rangle_{g_2} = \delta^{{}^{IJ}}\! \langle \psi_2 | \, S \, \{
\widetild{Q}_\I^\da, {Q}_\J^\db\}|\psi_1\rangle_\nul \ee First,
however, we need to verify whether the supersymmetry algebra is
indeed satisfied.

Let us first look at the interaction terms linear in $g_\two$.
Since the $SO(4)\times S0(4)$ rotation symmetry is purely
kinematical, it is clear that the rotation generators $J_{ij}$
should not receive any $g_\two$-corrections. Consistency of the
algebra to linear order in $g_\two$ thus requires that
\be \label{hcheck} \delta^{{}^{IJ}} \{ (Q_\nul)_{\I}^\da,
[(Q_\nul^\stt\! )_\J^\db,, \Sigma]\}\, = \, \delta^{\da\db} \, V_1
\ee with $V_1$ the order $g_\two$ interaction term of the
Hamiltonian.\footnote{Note however that, due to the $g_\two$
dependence of the inner product, the first order
$g_\two$-correction in the matrix element of $H$ in fact has two
contributions: $$\langle \psi_2| H_1 |\psi_1\rangle = \langle
\psi_2| (\Sigma H_0 + V_1) |\psi_1\rangle \, .$$} We will now show
that the above equation is indeed satisfied at the linearized
level in the fermions.

Explicit evaluation of the anti-commutator on the left-hand side
of  (\ref{hcheck}) gives (here we are omitting the spinor indices)
\be \{Q_\nul ,[Q_\nul^\stt\! ,\Sigma]\}
= \frac 1{2J^2} \sum_{m<n} \Sigma_{mn}\Bigl( \{Q_{\gamma_1}\!\! +
Q_{\gamma_2},Q_{\gamma_2}^\stt\! -Q_{\gamma_1}^\stt \}
-[Q_{\gamma_1}\!\! -Q_{\gamma_2},Q_{\gamma_2}^\stt\!
-Q_{\gamma_1}^\stt]\Bigr)
\ee where as before, we imagine that the operator is acting on a
state in the twisted sector $\gamma_1$; the permutation $\gamma_2$ is
obtained from $\gamma_1$ by applying the transposition $(mn)$.

The first term on the right-hand side is an anti-commutator of two
expressions of the same general form is in (\ref{expli}). It is
easy to see that this anti-commutator will produce an expression
of the form $\delta^{\da\db} V_1$. The second term, on the other
hand, is a {\it commutator} between anti-commuting quantities, and
is not proportional to $\delta^{\da\db}$. However, it is
necessarily quadratic in the fermion oscillators; moreover, one
can show it contains both fermionic annihilation and creation
operators (each acting at different locations), and therefore
gives a vanishing contribution when acting on purely bosonic
states in either direction. As stated before, we expect that this
term can be cancelled by adding higher order fermionic terms to
$Q$, without modifying the bosonic part of $H$.

It is not difficult to show that, when evaluated between bosonic
states, the supersymmetry algebra also closes to second order in
$g_\two$. Because the supercharges themselves are linear in
$g_\two$, this is sufficient. Hence we can use (\ref{hdiag}) to
define the bosonic matrix elements of $H$ to all orders in
$g_\two$. Inserting the original Ansatz (\ref{nsusy}) for the
supercharges, we obtain \be \label{all} \langle\, \psi_2| (
\delta^{\da\db} H\,+ J^{\da\db} ) | \, \psi_1 \rangle_{g_\two} =
\delta^{{}^{IJ}}
\langle \psi_2 | \, S\, ({Q}_\nul^{\gtt})_\I^{\da} S^{-1}
(Q_\nul^{\stt})_\J^\db S
|\,\psi_1\, \rangle_\nul   \; + \; (\da \leftrightarrow \db)
\ee Here we used that $Q^\gtt_\nul$ annihilates bosonic states; we
will continue to use this fact in the following. We will now
explicitly evaluate these matrix elements between the class of
states discussed earlier; we will work to leading order in
$\lambda'$ and second order in $g_\two$.

\newcommand{\sstt}{{{}^{<\!\! <}}}
\newcommand{\ggtt}{{{}^{>\!\! >}}}


\bigskip

\bigskip

\noindent {\sc Operator Mixing at Order $g_\two$}

To linear order in $g_\two$, the Hamiltonian has non-zero matrix
elements between single and double string states. In the gauge
theory, this corresponds to an operator mixing term between single
and double trace operators. We will find an exact match with the
gauge theory computations done recently in
\cite{bkpss}\cite{boston2}.

Expanding (\ref{all}) to linear order in $g_\two$ gives
\ba \langle \psi_2 | H_1 |\psi_1 \rangle \is  \langle \psi_2 | (
H_\nul \Sigma + \Sigma H_\nul
) |\psi_1\rangle
- 
\langle\psi_2| \widetild{Q}^\gtt_\nul \! \Sigma\,  Q_\nul^\stt\!
|\psi_1 \rangle.   
\ea It is straightforward to evaluate this amplitude between the
states $|1,p\rangle$ and $|2,k,y\rangle$ introduced in section 3.
We obtain \ba \label{triv} \langle 2, k, y | ( H_\nul \Sigma +
\Sigma H_\nul ) |1,p \rangle \! \is\! \lambda' (p^2 + k^2 /y^2) \,
C_{p ky}\, ,
\qquad \\[2.5mm]
\label{announce} \langle 2, k, y | \widetild{Q}^\gtt_\nul \,
\Sigma\, Q_\nul^\stt | 1, p\rangle \is  \lambda'\, (p k/y)  \,
C_{p ky}\, , \qquad
\ea
where $C_{pky}$ 
is the bare ``three point function'' (the matrix element of
$\Sigma$) given in (\ref{cijk}). The first result (\ref{triv}) is
immediate. The second result (\ref{announce}) can be understood
intuitively, by noting that after contracting the fermionic
oscillators in $Q_\nul^\stt$ and $Q_\nul^\gtt$, one is left with a
product of the form $\partial x^i \Sigma
\partial x^i$, which when evaluated gives a contribution
proportional to $pky$ times the bare three point function. We will
explicitly verify this intuition in Appendix A.

In a similar way, we can also evaluate the operator between
$|1,p,y\rangle$ and the other two string state $|2,y\rangle$. The
final result for both amplitudes \ba \langle 2 , k, y | \, H_1 \,
| 1, p\rangle \is \lambda'\, (p^2 +
k^2/ y^2 - p\, k/y ) C_{p\, ky} \\[2.5mm]
\langle 2, y | \, H_1 \, | 1, p\rangle \is \lambda'\, p^2  C_{p \,
y} \nonumber \ea reproduces the gauge theory result
\cite{bkpss}\cite{boston2}.

In itself, this match is not yet a conclusive indication that our
bit string theory reproduces the gauge theory amplitudes, because
the value of mixing matrix elements depends on a choice of basis.
It is an encouraging sign, however, that our present choice of
basis indeed matches with that of the gauge theory.\footnote{In
the concluding section, we will describe another choice of basis,
in which the inner product again becomes diagonal.} A more
conclusive verification of the correspondence requires going to
second order in $g_\two$.

\bigskip

\bigskip

\noindent{\sc Order $g_\two^2$ matrix element}

To compute the order $g_\two^2$ matrix element of $H$ between two
single string states, we first expand (\ref{all}) to second order.
We find
\be
\label{xyz}
\langle\, 1, q \, | \, H_2  | \, 1, p \,\rangle
= X - Y + Z
\ee
\vspace{-6mm}
with
\ba X  \is \frac 12 \langle  1, q | ( H_\nul
\Sigma^2 + \Sigma^2 \, H_\nul) |1, p\, \rangle 
\nonumber \\[2.5mm]
Y \is \frac 1{2} \, \langle 1, q |\,  \widetild{Q}_\nul^\gtt
\Sigma^2 \, Q_\nul^\stt
  |1, p \, \rangle 
\\[2.5mm]
Z \is 
\langle  1, q  | \, [\, \widetild{Q}_\nul^\gtt\!  , \Sigma ] [\,
\Sigma , Q_\nul^\stt]\, | 1, p \, \rangle \nonumber
\ea The first
two contributions $X$ and $Y$ have a similar structure as the
order $g_\two$ terms, except with $\Sigma$ replaced by $\Sigma^2$.
Again, $X$ is easily evaluated, while $Y$ can be obtained via a
similar calculation as before, with a similar result \ba
\label{xy} X \is  \lambda'(p^2 + q^2) A_{p\, q}\, , \qquad \qquad
\ \ Y \; = \;  \lambda' p\, q A_{p\, q} \, . \ea Here $A_{p\, q}$
is the matrix element of the splitting-and-joining interaction
term $\Sigma^2$, given in (\ref{adef}).

Finally we need to evaluate $Z$. This is done as follows:\\[-2mm]
\ba \label{zz}
Z \is \sum_i \Bigl( \langle 1, q | \, \Sigma | \, i
\rangle \langle \, i\, | [Q_\nul^\gtt\! , \Sigma] Q_\nul^\stt  \,
| 1,p\rangle \, - \, \langle 1, q |[ Q_\nul^\gtt\! , \Sigma ]
Q^\stt_\nul | \, i \, \rangle \langle
\, i\, |\Sigma | 1,p\rangle\Bigr)  \nonumber \\[2.5mm]
\is \sum_{k,y} \Bigl(p (k/y \!-\!p)\! -\! (k/y)(q\!-\! k/y) \Bigr)
C_{p\, ky}C_{q ky} \, - \, \sum_y p^2 C_{py}C_{qy}
\,\\[2.5mm]
\is  - (p^2 + q^2)A_{pq} + \sum_{k,y} {(k/y)^2}\;C_{pky}C_{q ky}
 \; \equiv \; {1\over 4\pi^2} B_{pq}
\nonumber
\ea Let us go over the individual steps of the above
calculation. In the first line, we used that
$[[Q^\gtt_\nul,\Sigma],\Sigma]\!=\!0$ and inserted a sum over a
complete set of intermediate two-string states. Next we evaluated
the matrix elements, using the previously obtained order $g_\two$
results (\ref{triv}) and (\ref{announce}). Finally, we used the
identities (\ref{adef}) and \cite{boston2} \ba
\sum_{k,y} (k/y) \, C_{p\, ky} C_{qky} \is (p+q) A_{p\, q},
\ea a relation that follows from the fact that the second line in
(\ref{zz}) must be a symmetric function of $p$ and $q$ (since $Z$
is a symmetric expression). The very last relation in (\ref{zz})
is the ``unitarity check'' of \cite{boston}\cite{boston2} (with
the correct sign!). Combined, the answers (\ref{xy}) and
(\ref{zz}), when put back into (\ref{xyz}), {\it exactly}
reproduce the gauge theory results of \cite{bkpss} and
\cite{boston2}. The $X$ and $Y$ contributions represent the
diagrams with nearest and semi-nearest neighbor contractions, and
the $Z$ term comprises all non-nearest neighbor interactions.

The knowledge of the second order matrix elements of $H$ and first
order mixing terms is sufficient to find the order $g^2$
corrections to the energy eigenvalues of two-impurity states. In
the gauge theory, these correspond to the leading $1/N$ correction
to their conformal dimensions.

\newsubsection{Conclusion}

We have presented an interacting string bit model and verified,
via a number of non-trivial tests, that it reproduces the
non-planar corrections to the gauge theory amplitudes. In
particular, for the special class of two-impurity operators, it
leads to the same $1/N$ corrections in the conformal dimensions as
reported in \cite{bkpss}\cite{boston2}\cite{gmr}.

Once this relation between the gauge theory and bit string theory
is established, however, we still have the freedom to choose a
different basis of states, that is, a different identification
between single and multi-string states and single and multi-trace
operators in the gauge theory. Given the form (\ref{ip}) of the
inner product, with $S$ as in (\ref{es2}), it is natural to define
a new basis of states via \be |\psi \rangle \rightarrow
|\widetilde{\psi}\rangle = e^{-g_\two \Sigma/2} |\psi\rangle \ee
Relative to this new basis of states, the inner product reduces to
the standard diagonal one, without any mixing terms between single
and multi-string states. The interacting supercharges in this
basis can be written as \be \label{nnnsusy}
Q = Q_\nul +
{g_\two\over 2} [Q^\stt_\nul\!\!-\!Q_\nul^\gtt,\Sigma]\,
.
\ee The strength of string interactions in this basis is in fact
$g_\two \sqrt{\lambda'}$, rather than $g_\two$: the string theory
at $\lambda'\! =\! 0$ is free. Note that the new interacting
supercharge (\ref{nnnsusy}) still has only a cubic interaction
term; it differs from the original Ansatz in \cite{bits} via a
(crucial) relative minus sign between $Q^\stt_\nul$ and
$Q_\nul^\gtt$, ensuring that (\ref{nnnsusy}) is hermitian.

We have shown that the superalgebra generated by these charges
closes at the linearized level in the fermions. There are two ways
in which one can try to improve our expression so that it
generates a closed algebra to all orders.  One approach would be
to try to modify the string bit theory by taking into account
that, in the gauge theory, the fermionic impurities carry half a
unit of R-charge, and in effect create or destroy half a string
bit. This modification may make it possible to write exact
supercharges, that are still linear in the fermions.

Alternatively, it should be possible to write non-linear fermionic
corrections to $Q$, designed to produce a closed algebra.
Presumably, this leads to an expression for the interaction vertex
similar to that of continuum light-cone string field theory
\cite{lcstring}\cite{sv}\cite{yj}, or its $SO(4)\times SO(4)$
invariant modification proposed in \cite{durham3}. In any case,
the string bit theory appears to provide a natural setting for
resolving some of the apparent discrepancies between the
perturbative gauge theory amplitudes
and the small $\lambda'$ limit
of string field theory amplitudes, 
which are most likely due to an unallowed interchange of limits
\cite{ksv}: to get the continuum string theory one should take the
large $J$ limit with {\it finite} $\lambda'$, rather than expand
around $\lambda'\!=\!0$ and then take the large $J$ limit. We
intend to return to these open question in a future work.

\bigskip

\newsubsection{Acknowledgements}

We thank David Berenstein, Dan Freedman, Matthew Headrick, Igor
Klebanov, Shiraz Minwalla, Lubos Motl, Sunil Mukhi, John Pearson,
Jan Plefka, Marc Spradlin and Matthias Staudacher for helpful
discussions. This research was supported by NSF grant 98-02484.


\bigskip
\bigskip

\appendix{}
In this appendix we derive eqn (\ref{announce}). Let $\gamma_1=(1~
2\ldots J)$ denote the single string sector and $\gamma_2=
(J~1\ldots J_1\!-1)(J_1 \ldots J-1)$ denote the double string
sector such that $\gamma_1 \circ\Sigma_{J_1J}\circ\gamma_2=1$. The
supersymmetry generators act on these sectors as (here we identify
the last site {\small $m=J$} with the 0-th site {\small $m=0$})
\ba Q^\stt_{\gamma_1}\! \is \! \lambda
\sum_{m=0}^{J-1}\bigg[(a_{m+1}^{i\ \dag}\! +a_{m+1}^i)-(a_m^{i\
\dag} \! +
a_m^i)\bigg]\gamma^i\beta_m^\dag\, ,
\\[2mm]
 \widetild{Q}^\gtt_{\gamma_2}\! \is \! \lambda \sum_{m=0}^{J-1}\beta_m\gamma^i
\Bigl[(a_{m+1}^{i\ \dag} \! +\! a_{m+1}^i)\! -\!
(a_m^i{}^\dag+a_m^i)\Bigr]
+\, (\beta_{{J_1\! -1}}\!\! -\!
\beta_{{J-1}})\gamma^i\bigg[(a_J^{i\ \dag}\! +\! a_J^i)
-(a_{J_1}^{i\ \dag}+ a_{J_1}^i)\bigg]\, .
\nonumber
\ea
It is convenient to further decompose $Q_{\gamma_1}^\stt$ into two
contributions: a term $Q_{\gamma_1}^\sgt$ containing all terms of
the form $a_i \gamma^i \beta^\dag$, and a second piece
$Q_{\gamma_1}^\sstt$ consisting of all contributions of the form
$a^\dagger_i \gamma^i \beta^\dag$. Similarly, we define
$Q_{\gamma_2}^\gst$ and $Q_{\gamma_2}^\ggtt$. We first consider
the matrix element $\langle \,  2| {Q}^{\gst}\Sigma\, Q^{\sgt} |1
\rangle$. We compute \ba Q^\sgt_{\gamma_1} O_J^p \is
-{\lambda\over J} \sum_{r,\, l=1}^J e^{\frac{2\pi i p(l-r)}{J}}
b^\dag_l (\beta_{r-1}^\dag - \beta_r^\dag)
 \\[3mm]
O^k_{J_1} Q^\gst_{\gamma_2}\is -{\lambda \over J_1}
\sum_{l'=0}^{J_1-1} \sum_{r'=1}^{J_1}\,  \Bigl(b_{l'}
e^{-\frac{2\pi i k l'}{J_1}} 
\Bigr)(\beta_{r'-1}-\beta_{r'})
e^{\frac{2\pi i k r'}{J_1}}  
 \nonumber \ea
\smallskip
Taking the inner product yields the sum
\smallskip
\ba 
{\lambda^2 \over JJ_1} \Bigl(\sum_{l=0}^{J_1-1}\! e^{2\pi i\, (
\frac{p}{J} -\frac{k}{J_1}) l}
\Bigr) \bigg(
\sum_{r,r'=1}^{J_1}\!\delta_{{r,r'}}+\!\sum_{r,r'=0}^{J_1\!
-1}\!\delta_{{r,r'}} -\! \sum_{r'=0}^{J_1\! -1}
\sum_{r=1}^{J_1}\!\delta_{{r'\! ,r -1}} -\! \sum_{r=0}^{J_1-1}
\sum_{r'=1}^{J_1}\! \delta_{{r,r'\! -1}}\! \bigg) e^{2\pi i \, (
\frac{k \, r'}{J_1} -\frac{ p\, r}{J})}\nonumber
\ea Performing the sum and using the operator state correspondence
(\ref{steen})-(\ref{stwee}), while keeping track of the action of
the centralizers, gives \ba \label{eerste} \langle 2,k , y|
{Q}^{\gst}\Sigma\, Q^{\sgt} |1, p \rangle \! \is \! {\lambda^2}\;
\langle 2,k,y| \Sigma |1,p \rangle \, \Bigl(1-e^{\frac{2\pi i
k}{J_1}}\Bigr) \Bigl(1-e^{-\frac{2\pi i p}{J}}\Bigr)
\\[2.5mm]
\langle 2,k,y| \Sigma |1,p \rangle \is {1\over J^2}
\sqrt{\mbox{\small $J-J_1$}\over \mbox{\small $JJ_1$}}
\sum_{l,r=0}^{J_1-1}e^{\frac{2\pi i (py -k)(l-r)}{J_1}} \, = \,
C_{pky}.\ea

The other matrix element $\langle 2,k,y| {Q}^{\ggtt}\Sigma
Q^{\sstt}|1,p\rangle$ turns out to be equal, at large $J$, to the
one we just computed, up to a ``normal ordering'' term \ba
\label{tweede} \langle 2, k,y | \bar{Q}^{\ggtt}\Sigma Q^{\sstt}|
1,p\rangle \is {\lambda^2 \over J^2}\sqrt{\mbox{\small
$J-J_1$}\over \mbox{\small $JJ_1$}}
\Bigl(\sum_{l=0}^{J_1-1}e^{-\frac{2\pi i (k-py)l}{J_1}}\Bigr) \,
\langle \gamma_2 |\Bigl(\sum_{r'=0}^{J_1-1}a_{r'} ^i e^{\frac{2\pi
i
kr'}{J_1}}\Bigr) \, \times \nonumber \\[2.5mm]& & \!\!\!\!\!
\bigg[\sum_{m=0}^{J-1} 
(a_{m+1}^i \! -a_m^i ) (a_{m+1}^{i\ \dag}\! -a_{m}^{i\ \dag})
\bigg]
\Bigl(\sum_{r=1}^J a_r^i{}^\dag e^{-\frac{2\pi i py
r}{J_1}}\Bigr)\, |\gamma_1
\rangle\nonumber\\[3mm]
&=&{2\lambda^2 J} \langle 2,k,y |\Sigma | 1,p \rangle \,
+ \, \langle 2,k,y | {Q}^{\gst}\Sigma \, Q^{\sgt}|1,p \rangle \,
.\eea The first term arises due to normal ordering and therefore
corresponds to a ``vacuum fluctuation''. In the end, this
contribution gets cancelled once we take into account the
infinitesimal readjustment of the vacuum induced by the presence
of the hopping terms in the Hamiltonian. After adding both
contributions (\ref{eerste}) and (\ref{tweede}), and taking the
limit $\frac k {J_1}, \frac p J <\!\! <1$, we arrive the announced
result (\ref{announce}).

\appendix{}
In this appendix we would like to present an explicit evaluation of the
action of the inner product $S$ on the string
vacuum
\ba
\langle \gamma_1|S |\gamma_1\rangle &=& \langle \gamma_1| e^{\frac{\Sigma_2}N}
|\gamma_1\rangle\nonumber\\
&=&\langle \gamma_1| \left(1+\sum_{h=1}^\infty\frac{1}{(2h)!
N^{2h}}\Sigma{}^{2h}\right) |\gamma_1\rangle \eea and show that it
indeed mirrors a gauge theory calculation \be \langle Tr
(Z^J)(0) Tr( Z^J) (x) \rangle = \left[1+ \sum_{h=1}^\infty
\frac{1}{(2h+1) !} \left(
\frac{g_2}2\right)^{2h}\right]\frac1{(4\pi^2 x^2)^J} \ee to order
$g_2{}^3$. Similar but more involved calculations are needed for
higher genera. Alternatively, a more general proof was given in
the main body of the paper.

To begin, let us evaluate the genus one contribution to the $S$-norm of
the vacuum.
Given a certain long string permutation $\gamma_1=(1 2 \dots J)$ the
action of $\Sigma_{ij}$ splits the string
into a two-string state $\gamma_2=(j~j +1 ~i+2\dots i-1)(1 2 \dots j-1 ~i~
i+1\dots J)$.
A further permutation $\Sigma_{kl}$ is needed to rejoin the strings, with
$j<k<i-1$ and $1<l<j,  i<l<J$.
Thus from
\ba
1\rightarrow 2\rightarrow 1 :~~~~ \frac{1}{2!N}\langle \gamma_1 |
\Sigma_2{}^2|\gamma_1\rangle &=& \frac{1}{2!N}
\sum_{i=1}^{J}\sum_{j=1}^i\sum_{k=i}^j\bigg(\sum_{l=1}^j +
\sum_{l=i}^J\bigg) ~1\nonumber\\
&=&
 \frac{1}{2!N}\sum_{i=1}^{J}\sum_{j=1}^i (i-j) (J-i+j)=
\frac{1}{3!}\left(\frac{J^2}{2N}\right)^2
\eea
one derives that the genus one contribution matches the genus one gauge
theory normalization of the
zero impurities BMN operators.

\figuur{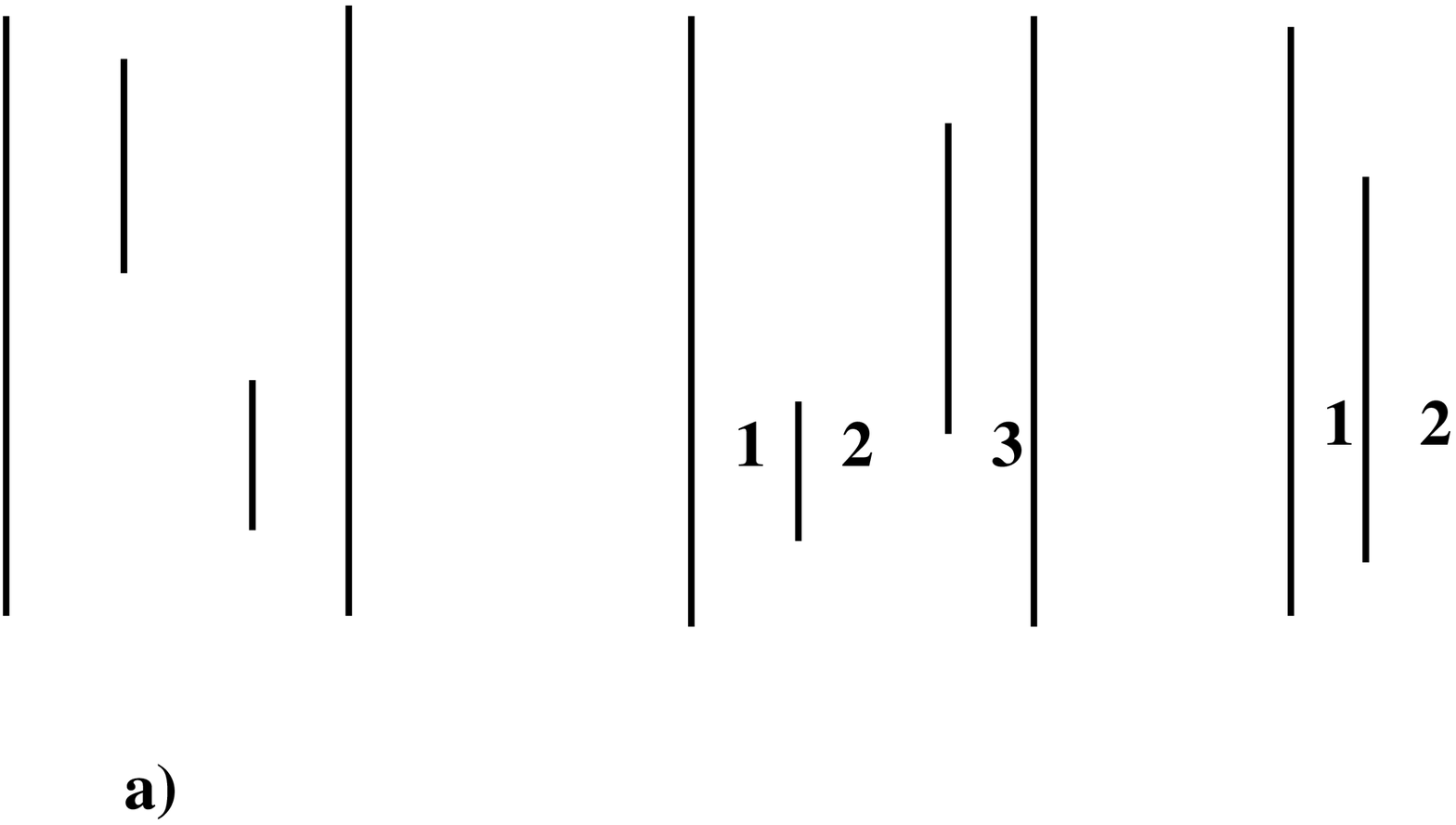}{10cm}{The light-cone string diagram of the
$1\rightarrow 2\rightarrow 1\rightarrow 2\rightarrow 1$ process
and b) the $1\rightarrow 2\rightarrow 3\rightarrow 2\rightarrow 1$
process}

The genus two calculation follows the same pattern: one first acts
with $\Sigma_{ij}$ to split the string. Next we can rejoin it with
$\Sigma_{kl}$, split it and join it once more. The net effect will
be the square of the previous splitting and joining computation :
\be 1\rightarrow 2\rightarrow1\rightarrow2\rightarrow 1 : ~~~~J^8
\frac{1}{3{}^2 4{}^2} \label{12}\ee
Or, we can decide to further
split the string after acting with $\Sigma_{ij}$, and join the
strings afterwards: $1\rightarrow 2\rightarrow 3\rightarrow
2\rightarrow 1$. Depending on the position of the $k, l$ indices
with respect to $i,j$ we distinguish four cases: i) $j<l<k<i$, ii)
$l<k<j<i$, iii) $l<j<i<k$, and iv) $ j<i<l<k$. In each case the
three strings obtained after the action of
$\Sigma_{kl}\Sigma_{ij}$ can be joined in three different ways by
acting further with two consecutive transpositions. Implicitly
performing the summation over the position of the indices of the
latter two transpositions one obtains \ba
&&\!\!\!\!\!\!\!\!\!\!\!\!\!\!\!\!\!\!\!\!\!\!\!\!\! 1\rightarrow
2\rightarrow 3\rightarrow 2\rightarrow 1:~~~~ i)+ ii)+
iii)+iv)=\frac{1}{630}J^8+\frac{1}{840}J^8+\frac{1}{630}J^8+\frac{1}{840}J^8=\frac{1}{180}J^8
\nonumber\\
\label{123}\\
i)&=& \sum_{i=1}^J\sum_{j=1}^i\sum_{k=j}^i\sum_{l=j}^k\bigg[
(l\! - \!j\! + \!i\! - \!k)(k\! - \!l)(i\! - \!j)(j\! + \!J\! - \!i)\nonumber\\
&\! + \!&(k\! - \!l)(j\! + \!J\! - \!i)(k\! - \!l\! + \!j\! + \!J\! - \!i)(l\! - \!j\! + \!i\! - \!k)\! + \!(l\! - \!j\! + \!i\! - \!k)(j\! + \!J\! - \!i)(l\! + \!J\! - \!k)(k\! - \!l)\bigg]\nonumber\\
ii)&=&\sum_{i=1}^J\sum_{j=1}^i\sum_{k=1}^j\sum_{l=1}^k\bigg[(i\! -
\!j)(k\! - \!l)(i\! - \!j\! + \!k\! - \!l)(l\! + \!j\! - \!k\! +
\!J\! - \!i)
\nonumber\\
&\! + \!&(i\! - \!j)(l\! + \!j\! - \!k\! + \!J\! - \!i)(l\! + \!J\! - \!k)(k\! - \!l)\! + \!(k\! - \!l)(l\! + \!j\! - \!k\! + \!J\! - \!i)(j\! + \!J\! - \!i)(i\! - \!j)\bigg]\nonumber\\
iii)&=&\sum_{i=1}^J\sum_{j=1}^i\sum_{k=i}^J\sum_{l=1}^j\bigg[(i\!
- \!j)(k\! - \!i\! + \!j\! - \!l)(k\! - \!l)(l\! + \!J\! - \!k)
\nonumber\\
&\! + \!&(i\! - \!j)(l\! + \!J\! - \!k)(l\! + \!J\! - \!k\! + \!i\! - \!j)(j\! - \!l\! + \!k\! - \!i)\! + \!(j\! - \!l\! + \!k\! - \!i)(l\! + \!J\! - \!k)(j\! + \!J\! - \!i)(i\! - \!j)\bigg]\nonumber\\
iv)&=&\sum_{i=1}^J\sum_{j=1}^i\sum_{k=i}^J\sum_{l=i}^k\bigg[( i\!
- \!j)(k\! - \!l)(i\! - \!j\! + \!k\! - \!l)(j\! + \!l\! - \!i\! +
\!J\! - \!k) \nonumber\\&\! + \!&(i\! - \!j)(j\! + \!l\! - \!i\! +
\!J\! - \!k)(l\! + \!J\! - \!k)(k\! - \!l)\! + \!(k\! - \!l)(j\! +
\!l\! - \!i\! + \!J\! - \!k)(j\! + \!J\! - \!i)(i\! - \!j)\bigg]
\eea Putting everything (\ref{12},\ref{123}) together, the genus
two contribution to the vacuum $S$-norm is \ba
\frac{1}{4!N^2}\langle \gamma_1 | \Sigma_2^{4}|\gamma_1\rangle =
\frac{J^8}{4!N^2}\left(\frac{1}{180}+\frac{1}{3{}^2 4{}^2}\right)
=\frac1{5!}\left(\frac{J^2}{2N}\right)^4 \eea once more in perfect
agreement with the gauge theory calculation.



\begin{thebibliography}{30}
\parskip-2pt

\bibitem{bmn}{ D.~Berenstein, J.~M.~Maldacena and H.~Nastase,
``Strings in flat space and pp waves from N = 4 super Yang
Mills,'' JHEP {\bf 0204}, 013 (2002) [arXiv:hep-th/0202021].
}
\bibitem{bla}{ C.~M.~Hull, ``Killing Spinors And Exact Plane Wave
Solutions Of Extended Supergravity,'' Phys.\ Rev.\ D {\bf 30}, 334
(1984).
M.~Blau, J.~Figueroa-O'Farrill, C.~Hull and G.~Papadopoulos, ``A
new maximally supersymmetric background of IIB superstring
theory,'' JHEP {\bf 0201}, 047 (2002) [arXiv:hep-th/0110242].
;M.~Blau, J.~Figueroa-O'Farrill, C.~Hull and G.~Papadopoulos,
``Penrose limits and maximal supersymmetry,'' Class.\ Quant.\
Grav.\  {\bf 19}, L87 (2002) [arXiv:hep-th/0201081].
}
\bibitem{bits}{ H.~Verlinde, ``Bits, Matrices and 1/N,''
arXiv:hep-th/0206059.
}
\bibitem{thornbits}
{R.~Giles and C.~B.~Thorn,
``A Lattice Approach To String Theory,''
Phys.\ Rev.\ D {\bf 16}, 366 (1977);
C.~B.~Thorn, {
``Supersymmetric quantum mechanics for
string-bits,''} Phys.\ Rev.\ D {\bf 56}, 6619 (1997)
[arXiv:hep-th/9707048];
A related approach to large $N$ gauge theory has been formulated
in:
C.~B.~Thorn,
``A Fock Space Description Of The 1/N-C Expansion Of Quantum Chromodynamics,''
Phys.\ Rev.\ D {\bf 20}, 1435 (1979).}

\bibitem{bkpss} {N.~Beisert, C.~Kristjansen, J.~Plefka,
G.~W.~Semenoff and M.~Staudacher, ``BMN Correlators and Operator
Mixing in N=4 Super Yang-Mills Theory,'' arXiv:hep-th/0208178.
}
\bibitem{boston2}{
N.~R.~Constable, D.~Z.~Freedman, M.~Headrick and S.~Minwalla,
``Operator mixing and the BMN correspondence,''
arXiv:hep-th/0209002.}
\bibitem{gmr}{ D.~J.~Gross, A.~Mikhailov and R.~ Roiban. ``A
calculation of the plane wave string Hamiltonian from N=4
super-Yang-Mills theory,'' archiv:hep-th/0208231}
\bibitem{boston}{ N.~R.~Constable, D.~Z.~Freedman, M.~Headrick,
S.~Minwalla, L.~Motl, A.~Postnikov and W.~Skiba, ``PP-wave string
interactions from perturbative Yang-Mills theory,'' JHEP {\bf
0207}, 017 (2002) [arXiv:hep-th/0205089].
}
\bibitem{kpss}{ C.~Kristjansen, J.~Plefka, G.~W.~Semenoff and
M.~Staudacher, ``A new double-scaling limit of N = 4 super
Yang-Mills theory and PP-wave  strings,'' arXiv:hep-th/0205033.
}
\bibitem{bn}{ D.~Berenstein and H.~Nastase, ``On lightcone
string field theory from super Yang-Mills and holography,''
arXiv:hep-th/0205048.
}
\bibitem{durham}{ C.~S.~Chu, V.~V.~Khoze and G.~Travaglini, ``Three-point
functions in N = 4 Yang-Mills theory and pp-waves,'' JHEP {\bf
0206}, 011 (2002) [arXiv:hep-th/0206005]; ``pp-wave string
interactions from n-point correlators of BMN operators,''
arXiv:hep-th/0206167.
}
\bibitem{huang}{ M.~x.~Huang, ``String interactions in pp-wave from N
= 4 super Yang Mills,'' arXiv:hep-th/0206248;
``Three point functions of N = 4 super Yang Mills from light cone
string  field theory in pp-wave,'' Phys.\ Lett.\ B {\bf 542}, 255
(2002) [arXiv:hep-th/0205311].
}
\bibitem{bianchi}{ M.~Bianchi, B.~Eden, G.~Rossi and Y.~S.~Stanev,
``On operator mixing in N = 4 SYM,'' arXiv:hep-th/0205321.
}
\bibitem{mets}{ R.~R.~Metsaev, ``Type IIB Green-Schwarz superstring
in plane wave Ramond-Ramond  background,'' Nucl.\ Phys.\ B {\bf
625}, 70 (2002) [arXiv:hep-th/0112044].
R.~R.~Metsaev and A.~A.~Tseytlin, ``Exactly solvable model of
superstring in plane wave Ramond-Ramond  background,'' Phys.\
Rev.\ D {\bf 65}, 126004 (2002) [arXiv:hep-th/0202109].
}
\bibitem{zhou}
J.~G.~Zhou, ``PP-wave string interactions from string bit model,''
arXiv:hep-th/0208232.
\bibitem{lcstring}{ M.~B.~Green and J.~H.~Schwarz, ``Superstring
Interactions,'' Nucl.\ Phys.\ B {\bf 218}, 43 (1983).
}
\bibitem{sv}{ M.~Spradlin and A.~Volovich, ``Superstring interactions
in a pp-wave background,'' arXiv:hep-th/0204146;
 ``Superstring interactions
in a pp-wave background II,'' arXiv:hep-th/0206073.
}
\bibitem{yj}{
Y.~j.~Kiem, Y.~b.~Kim, S.~m.~Lee and J.~m.~Park, ``pp-wave /
Yang-Mills correspondence: An explicit check,''
arXiv:hep-th/0205279;
P.~Lee, S.~Moriyama and J.~w.~Park, ``Cubic interactions in
pp-wave light cone string field theory,'' arXiv:hep-th/0206065}
\bibitem{durham3}
C.~S.~Chu, V.~V.~Khoze, M.~Petrini, R.~Russo and A.~Tanzini, ``A
note on string interaction on the pp-wave background,''
arXiv:hep-th/0208148.
\bibitem{ksv}
{I.~R.~Klebanov, M.~Spradlin and A.~Volovich, ``New effects in
gauge theory from pp-wave superstrings,'' arXiv:hep-th/0206221.
}





\end{thebibliography}
\end{document}